\documentstyle[twocolumn,eqsecnum,prb,aps]{revtex}

\newcommand{\la}{\langle}
\newcommand{\ra}{\rangle}

\begin{document}   
 
\draft  

\title{Electromagnetic characteristics 
of bilayer quantum Hall systems\\ in the presence of interlayer
coherence and tunneling }
 \author{K. Shizuya}
  \address{Yukawa Institute for Theoretical Physics\\
 Kyoto University,~Kyoto 606-8502,~Japan }

\maketitle

\begin{abstract} 
The electromagnetic characteristics of bilayer quantum Hall systems in the
presence of interlayer coherence and tunneling are studied by means of a
pseudospin-texture effective theory and an algebraic framework of the
single-mode approximation, with emphasis on clarifying the nature of the
low-lying neutral collective mode responsible for interlayer tunneling
phenomena.  A long-wavelength effective theory, consisting of the
collective mode as well as the cyclotron modes, is constructed.
It is seen explicitly from the electromagnetic response that gauge
invariance is kept exact, this implying, in particular, the absence of
the Meissner effect in bilayer systems.   Special emphasis is placed on
exploring the advantage of looking into quantum Hall systems through
their response; in particular, subtleties inherent to the standard
Chern-Simons theories are critically examined.
\end{abstract}

\pacs{73.43.Lp, 73.21.Ac}

\section{Introduction}

The Chern-Simons (CS) theories, both bosonic~\cite{ZHK,LZ,Z} and
fermionic,~\cite{BW,LF,HLR}  realize the composite-boson and
composite-fermion descriptions~\cite{GM} of the fractional quantum Hall
effect~\cite{TSG,L} (FQHE) and have been successful in
describing various features of the FQHE. 
They, however, have some subtle limitations as well.~\cite{MZ}   
In particular, when applied to bilayer systems,
they differ significantly in collective-excitation spectrum from the 
magneto-roton theory of Girvin, MacDonald and Platzman,~\cite{GMP} based
on the single-mode approximation (SMA).

The quantum Hall effect exhibits a variety of physics for bilayer 
(and multilayer) 
systems.~\cite{BIH,Chak,MPB,WZdlayer,EI,RR,Yang,LFdlayer,Moon,SES,Murphy} 
In a previous paper~\cite{KSdl} we studied within the SMA theory the
electromagnetic characteristics of bilayer systems in the absence of
interlayer coherence and derived a long-wavelength effective theory
that properly  embodies the SMA spectrum of collective excitations.
The effective theory was constructed from the electromagnetic response
of the systems through functional bosonization,~\cite{FS} without
referring to the composite bosons or composite fermions.  Thereby the
relation between the SMA theory and the CS theories was examined.

The purpose of the present paper is to extend the program of
looking into quantum Hall systems through their response  
to situations of particular interest, bilayer systems in the presence of
interlayer coherence as well as tunneling, where 
phenomena such as a crossover between the tunneling and
coherence regimes~\cite{MPB,Murphy} and Josephson-like
effects~\cite{WZdlayer,EI,SEPW,BR} attract attention.  We study
the electromagnetic characteristics of bilayer systems by means of 
(i) a pseudospin-texture effective theory and (ii) an algebraic
framework of the single-mode approximation, with essentially the same
results. Our analysis shows that proper account of Landau-level
projection is indispensable for deriving a low-energy effective theory
of gauge-invariant form.  The presence of interlayer coherence
modifies even the leading long-wavelength features of the bilayer
systems and we critically examine the CS approach to clarify its
validity and limitations.

In Sec.~II we consider the projection of a bilayer system into the Landau
levels. We study the electromagnetic characteristics of the bilayer system 
in Secs.~III and IV. In Sec.~V we comment on the CS approach.
Section~VI is devoted to a summary and discussion.

\section{bilayer systems} 

Consider a bilayer system with  average electron densities 
$\rho_{\rm av}^{(\alpha)}=(\rho_{\rm av}^{(1)},\rho_{\rm av}^{(2)})$ 
in the upper $(\alpha =1)$ and lower $(\alpha =2)$ layers.  
The two layers, each extending in the ${\bf x}=(x_{1},x_{2})$ plane, are
taken to be situated at position
$z^{(1)}= z_{c} + {1\over{2}}\, d$ and $z^{(2)}= z_{c} - {1\over{2}}\, d$
with separation $d$ in the vertical $(z)$ direction.
The system is placed in a common strong perpendicular magnetic field
$B_{z}=B>0$. We suppose that the electron fields 
$\psi^{(\alpha)}$ in each layer are fully spin polarized
and assemble them into a pseudospin~\cite{Moon} doublet spinor
$\Psi=(\psi^{(1)},\psi^{(2)})^{\rm tr}$.  
Our task in this paper is to study how the system responds to weak
electromagnetic potentials $A_{\mu}(x; z)$ and $A_{z}(x;z)$ in three
space. 
[We  suppose $\mu$ runs over
$(0,1,2)$ or $(t,x_{1},x_{2})$ and denote
$A_{k}=(A_{1},A_{2})={\bf A}$ and $x=(t,{\bf x})$ for short.]  We
thus write the one-body Lagrangian in the form
\begin{eqnarray}
&&L_{1} =  \int d^{2}{\bf x}\,  \Psi^{\dag}\,
(i\partial_{t}- {\cal H})\, \Psi,
\label{Selectron}\\
&&{\cal H} = {1\over{2M}}({\bf p} \!+ \! {\bf A}^{\!B}\!
+ \! {\bf A}^{\!+}\! + {\bf A}^{\!-}\sigma_{3})^{2} 
+ A_{0}^{\!+} +A_{0}^{\!-} \sigma_{3}, 
\label{Hone}
\end{eqnarray}
where
$A_{\mu}^{\pm}(x)= {1\over{2}}\,\{A_{\mu}(x;z^{(1)}) \pm
A_{\mu}(x;z^{(2)})\}$
in terms of the potentials acting on each layer, or explicitly,
\begin{eqnarray}
A^{+}_{\mu}(x) &=&  A_{\mu}(x;z_{c}) +\cdots,\nonumber\\
A^{-}_{\mu}(x) &=& (d/2)\,\partial_{z_{c}} A_{\mu}(x;z_{c})
+\cdots;
\end{eqnarray}
${\bf A}^{\!B}= eB\,(-x_{2},0)$ supplies a uniform magnetic field $B$; the
electric charge $e>0$ has been suppressed by rescaling
$eA_{\mu}\rightarrow A_{\mu}$;
[For conciseness, we shall write $\psi^{(\alpha)}(x) =
\psi^{(\alpha)}({\bf x}, z^{(\alpha)},t)$, etc., and suppress reference
to the $z$ coordinate or
$z_{c}$ unless necessary.]  Let us denote the number density and pseudospin
densities as
\begin{eqnarray}
\big\{\rho (x)\,, S^{a}(x)\big\}= \Psi^{\dag}(x)\,
\big\{1\, , {\scriptstyle{1\over{2}}}\,\sigma_{a}\big\}\,\Psi (x)
\end{eqnarray}
with the Pauli matrices $\sigma_{a}\ (a=1,2,3)$.  
The $A_{0}^{+}$ coupled to $\rho=\rho^{(1)} + \rho^{(2)}$ probes
in-phase density fluctuations of the two layers while $A_{0}^{-}$ coupled to 
$S^{3}={1\over{2}}\,(\rho^{(1)} -\rho^{(2)})$ probes the out-of-phase
density fluctuations.

The electrons in the two layers are coupled through the intralayer and
interlayer Coulomb potentials $V^{11}_{\bf p}=V^{22}_{\bf p}$ and 
$V^{12}_{\bf p} = V^{21}_{\bf p}$, respectively; 
$V^{11}_{\bf p}= e^{2}/(2\epsilon\, |{\bf p}|)$ and 
$V^{12}_{\bf p} =e^{-d |{\bf p}|}V^{11}_{\bf p}$ with $\epsilon$
being the dielectric constant  of the substrate.
The pseudospin structure of the Coulomb interaction is made manifest
by rewriting it as 
\begin{eqnarray}
H^{\rm C}
&=&{1\over{2}}\, \sum_{\bf p} \Big( V^{+}_{\bf p}\,
\rho_{\bf- p}\,\rho_{\bf p} +
4\, V^{-}_{\bf p}\, S^{3}_{\bf- p}\, S^{3}_{\bf p} \Big) ,
\label{HCoul} \\
V^{\pm}_{\bf p}&=&{1\over{2}}(1\pm e^{-d|{\bf p}|})V^{11}_{\bf p},
\end{eqnarray}
where $\rho_{\bf p}$ and $S^{a}_{\bf p}$ stand for the Fourier
transforms of $\rho(x)$ and $S^{a}(x)$ with obvious time dependence
suppressed.

Note here that the electromagnetic gauge transformations in three space
induce two sets of intralayer gauge transformations 
$A_{\mu}(x;z^{(\alpha)})\rightarrow A_{\mu}(x;z^{(\alpha)})
+\partial_{\mu}\, \theta^{(\alpha)} (x)$ and 
$\psi^{(\alpha)}(x)\rightarrow
e^{-i\theta^{(\alpha)}(x)}\psi^{(\alpha)}(x)$ 
with $\theta^{(\alpha)}(x) \equiv \theta (x;z^{(\alpha)})$, which can be
regarded as totally independent (for $d\not=0$) since $\theta (x;z)$
may have arbitrary dependence on $z$.  
The transformation laws read 
$\delta A_{\mu}^{\pm}(x) = \partial_{\mu}\theta^{\pm}(x)$ in terms of
$\theta^{\pm}= {1\over{2}}\,\{\theta^{(1)} \pm \theta^{(2)} \}$.  
Thus, for bilayer systems electromagnetic gauge invariance turns
into two separate U(1) gauge symmetries,
U(1)$^{\rm em} \supset$ U(1)$^{+} \times$ U(1)$^{-}$.
[We refer to this U(1)$^{-}$ as \lq\lq interlayer" gauge invariance
below. Note that it disappears in the $d \rightarrow 0$
limit.]

The tunneling phenomena must respect electromagnetic gauge invariance. 
A naive choice of interlayer coupling 
$S^{1} + iS^{2} = \psi^{(1)\dag}\,\psi^{(2)}$ 
should be promoted to a gauge-invariant form~\cite{Etwo}
\begin{eqnarray}
H^{\rm tun} &=& -\triangle_{SAS} \int d^{2}{\bf x}\,
{1\over{2}}\,\Big\{ \psi^{(1)\dag}\,e^{-i\Gamma_{z}}\,\psi^{(2)} +
{\rm h.c.}  \Big\}, 
\end{eqnarray}
with the line integral 
\begin{eqnarray}
\Gamma_{z}(x) =\int_{z^{(2)}}^{z^{(1)}} dz\, A_{z}(x;z) = d\,
A_{z}(x;z_{c}) + \cdots,
\end{eqnarray}
connecting the two layers for each ${\bf x}$. 
Here $\Gamma_{z}$ has the transformation law 
$\delta \Gamma_{z} = 2\theta^{-}(x)$.
The coupling strength $\triangle_{\rm SAS}$ characterizes the energy gap
between the symmetric and antisymmetric states.

It is possible to gauge away  $\Gamma_{z}$ by setting
$\Gamma'_{z}=\Gamma_{z} +2 \theta^{-}=0$ so that the transformed fields 
\begin{eqnarray} 
\psi'^{(1)}(x)&=& e^{i{1\over{2}}\, \Gamma_{z}}\, \psi^{(1)}(x),\ 
\nonumber\\
\psi'^{(2)}(x) &=& e^{-i{1\over{2}}\, \Gamma_{z}}\, \psi^{(2)}(x),
\nonumber\\
A'^{-}_{\mu}(x) &=& A^{-}_{\mu}(x) - {\scriptstyle {1\over{2}}}\, 
\partial_{\mu}\Gamma_{z}
\nonumber\\
&=& (d/2)\, (\partial_{z}A_{\mu} - \partial_{\mu}A_{z}) + \cdots,
\label{transformedfield}
\end{eqnarray}  
and ${A'}_{\mu}^{+}(x) =A_{\mu}^{+}(x)$ are inert under U(1)$^{-}$ gauge
transformations.
The $\psi'^{(\alpha)}(x)$ stand for the electron fields \lq\lq projected"
to the common plane $z=z_{c}$ and undergo only the U(1)$^{+}$ gauge
transformations.
Note that $A'^{-}_{\mu}(x)$ is gauge
invariant and actually denotes  a vertical electric field
$A'^{-}_{0}\approx (d/2)\, E_{z}$ and in-plane magnetic fields
$(A'^{-}_{1}, A'^{-}_{2}) \approx (d/2)\, (B_{2}, -B_{1})$. 

In view of this structure it is advantageous to restart with the
Lagrangian written in terms of these $\psi'^{(\alpha)}$ and
$A'^{\pm}_{\mu}$, and recover the effect of $\Gamma_{z}$ at the very
end.  The interlayer gauge invariance is thereby kept exact. 
Accordingly we shall from now on regard 
$\psi^{(\alpha)}$ and $A^{\pm}_{\mu}$ as denoting the
transformed fields $\psi'^{(\alpha)}$ and $A'^{\pm}_{\mu}$.

In addition, it is rather natural and convenient to combine the
one-body Lagrangian $L_{1}$ and $H^{\rm tun}$
into a formally  U(1)$\times$SU(2) gauge symmetric form 
by setting 
\begin{eqnarray} 
A_{\mu}^{-}\,\sigma_{3}
\rightarrow A_{\mu}^{a}\,{\scriptstyle {1\over{2}}}\,\sigma_{a}
\end{eqnarray}  
in ${\cal H}$ of Eq.~(\ref{Hone}) and by identifying the SU(2) gauge field
$A_{\mu}^{a} \equiv  (A_{\mu}^{1},A_{\mu}^{2},A_{\mu}^{3})$ with 
\begin{eqnarray} 
A^{1}_{\mu}=-\triangle_{SAS}\, \delta_{\mu 0},\ 
A_{\mu}^{2} =0,\ 
A_{\mu}^{3} = 2{A'}_{\mu}^{-}.
\label{Asutwo}
\end{eqnarray}  
This SU(2) gauge symmetry, of course, is only superficial.
The system has a global pseudospin SU(2) symmetry in the ideal limit 
$\triangle_{SAS}\rightarrow 0$  and 
$V^{-}_{\bf p}\rightarrow 0$ with ${A'}_{\mu}^{-}=0$; it gets broken to
U(1) either for $\triangle_{SAS} = 0$ or for $V^{-}_{\bf p} =0$.

The \lq\lq interlayer" gauge invariance has to
do with interlayer out-of-phase U(1) rotations induced by the $z$
variation of $\theta (x;z)$, i.e.,
$\theta^{-}\propto \partial_{z}\theta (x;z)$.  
They are thus {\em distinct} from global U(1) rotations (with constant
$\theta^{-}$) about the $S^{3}$ axis, which have to do with charge
conservation.
As a result, the tunneling interaction $H^{\rm tun} \propto
S'^{1}(x)$ defined in terms of $\psi'^{(a)}$ is gauge-invariant but
transforms covariantly (i.e., breaks invariance) under global U(1)$^{-}$
rotations. (This in turn implies that there is no loss of
generality in choosing $H^{\rm tun} \propto S'^{1}$.)

Let us now project our system onto the Landau levels.  
Let $|N\rangle = |n,y_{0}\rangle$ denote the Landau levels of a freely orbiting
electron of energy $\omega_{c} (n+{\scriptstyle {1\over2}})$ with 
$n = 0,1,2,\cdots$, and $y_{0}=\ell^{2}\,p_{x}$, where
$\omega_{c} \equiv eB/M$ and ${\ell}\equiv 1/\sqrt{eB}$; 
we frequently set $\ell \rightarrow 1$ below.  
We first pass into $N=(n,y_{0})$ space via a unitary transformation
$\Psi ({\bf x},t)=\sum_{N} \langle {\bf x}|N\rangle\, \Phi_{n}(y_{0},t)$
and, by a subsequent unitary transformation 
$\Phi_{n}(y_{0},t) \rightarrow \Psi_{m}(y_{0},t)$, 
make the one-body Hamiltonian diagonal in level indices; the relevant
transformation is constructed in powers of
$A^{+}_{\mu}$ and $A^{a}_{\mu}$. The resulting projected Hamiltonian is
an operator in 
${\bf r} \equiv (r_{1},r_{2})
= (i\ell^{2} \partial/\partial y_{0}, y_{0})$
with uncertainty $[r_{1}, r_{2}]=i\ell^{2}$.  
Such a systematic procedure of projection, developed
earlier,~\cite{KSdl,KS} is readily adapted to the present SU(2) case.   
As a matter of fact, for $\triangle_{SAS}\ll\omega_{c}$ the result is
essentially the same as in the single-layer case.

Let us focus on the lowest Landau level $n=0$ in a strong magnetic field.
The projected one-body Hamiltonian to $O(A^{2})$ reads 
$\bar{H}^{\rm cyc} +\bar{H}^{\rm em} + \bar{H}^{\rm tun}$ with
\begin{eqnarray}
\bar{H}^{\rm cyc} &=& \sum_{\alpha=1}^{2}
\sum_{\bf p}\Big\{{\omega_{c}\over{2}}\, \delta_{\bf p,0}
+{\cal U}^{(\alpha)}_{\bf p}\Big\} \,
\bar{\rho}_{\bf -p}^{(\alpha)} ,\nonumber\\
\bar{H}^{\rm em} &=& \sum_{\bf p}\Big\{ \chi^{+}_{\bf p} \,
\bar{\rho}_{\bf -p} +2 \chi^{-}_{\bf p}\, \bar{S}^{3}_{\bf -p}\Big\} ,
\nonumber\\
\bar{H}^{\rm tun} &=& - \triangle_{SAS}\, \bar{S}^{1}_{\bf p=0} ,
\end{eqnarray}
with $\chi_{\bf p}^{\pm} =(A_{0}^{\pm})_{\bf p} +
(1/2M)({A}_{12}^{\pm})_{\bf p}$ and 
$A_{12}^{\pm} = \partial_{1}A_{2}^{\pm} - \partial_{2}A_{1}^{\pm}$; 
$(A_{\mu}^{\pm})_{\bf p}$ stands for the Fourier transform of
$A_{\mu}^{\pm}(x)$.
Here the projected charges 
$\bar{\rho}_{\bf p}= \bar{\rho}^{(1)}_{\bf p} +
\bar{\rho}^{(2)}_{\bf p}$, 
$\bar{S}^{3}_{\bf p}= 
{1\over{2}}\,(\bar{\rho}^{(1)}_{\bf p} -\bar{\rho}^{(2)}_{\bf p})$,
etc., are defined by  
\begin{eqnarray} 
 \bar{\rho}_{\bf p}&\equiv&
\int dy_{0}\ \Psi_{0}^{\dag}(y_{0},t)\,
e^{-{1\over{4}}\, {\bf p}^{2}}\, e^{-i{\bf p\cdot r}}\,
\Psi_{0}(y_{0},t), \\
 \bar{S}^{a}_{\bf p}&\equiv&
\int dy_{0}\ \Psi_{0}^{\dag}(y_{0},t)\,
e^{-{1\over{4}}\, {\bf p}^{2}}\, e^{-i{\bf p\cdot r}}\,
{\sigma_{a}\over{2}}\,\Psi_{0}(y_{0},t),
\end{eqnarray}
where the two-spinor $\Psi_{0}$, defining the true lowest Landau level,
obeys the canonical commutation relation
$\{\Psi_{0}(y_{0},t), {\Psi_{0}}^{\dag}(y'_{0},t)\} 
= \delta (y_{0}-y'_{0})$.
The ${\cal U}^{(\alpha)}_{\bf p}$ denote the contributions quadratic
in $A_{\mu}^{(\alpha)}$, and are given (for ${\bf p}=0$) by 
${\cal U}^{(\alpha)}_{\bf p=0}= 
\int d^{2}{\bf x}\, {\cal U}^{(\alpha)}(x)$  with
\begin{eqnarray}
{\cal U}^{(\alpha)}
&=& {1\over{2}} A^{(\alpha)}_{\mu} D \epsilon^{\mu \nu
\rho}\partial_{\nu}A^{(\alpha)}_{\rho} \!
-{1\over{2\omega_{c}}} A^{(\alpha)}_{k0} D A^{(\alpha)}_{k0} 
+\cdots, \label{Utwo}
\end{eqnarray}
where $D = \omega_{c}^{2}/(\omega_{c}^{2} + \partial_{t}^{2} )$;  
$A_{\mu \nu}= \partial_{\mu}A_{\nu}- \partial_{\nu}A_{\mu}$, and
$\epsilon^{\mu\nu\rho}$ is a totally-antisymmetric
tensor with $\epsilon^{012} =1$.  Here we have retained terms to 
$O(\nabla^{2}/\omega_{c})$;
see Ref.~\onlinecite{KSdl} for an expression exact to all powers of
$\partial_{k}$.

The charges $(\bar{\rho}_{\bf p},\bar{S}^{a}_{\bf p})$ obey an
SU(2)$\times W_{\infty}$ algebra~\cite{Moon}
\begin{eqnarray}
&&[\bar{\rho}_{\bf p}, \bar{\rho}_{\bf k}]
= -2is(p,k) \, \bar{\rho}_{\bf p+k} , \ \ \ 
[\bar{\rho}_{\bf p}, \bar{S}^{a}_{\bf k}]
= -2is(p,k) \, \bar{S}^{a}_{\bf p+k}, \nonumber\\
&&[\bar{S}^{a}_{\bf p},\bar{S}^{b}_{\bf k}]= 
c(p,k)\,i\epsilon^{abc}\, \bar{S}^{c}_{\bf p+k}
-\delta^{ab}\, {i\over{2}}\,  s(p,k)\,\bar{\rho}_{\bf p+k},
\label{chargealgebra}
\end{eqnarray}
where 
\begin{eqnarray}
s(p,k) &=& \sin \Big({{\bf p}\!\times\! {\bf k}\over{2}} \Big)\ 
e^{{1\over{2}}\, {\bf p\cdot k}};
\end{eqnarray}
$c(p,k)$ is given by $s(p,k)$ with  $\sin \rightarrow \cos$.
It is important to note here that the projected charges themselves, 
$(\rho_{00}^{(\alpha)})_{\bf p}= \bar{\rho}^{(\alpha)}_{\bf p} +
\triangle \bar{\rho}^{(\alpha)}_{\bf p}$, 
differ slightly~\cite{KSdl} from $\bar{\rho}^{(\alpha)}_{\bf p}$ by 
$A^{(\alpha)}_{\mu}$-dependent corrections 
$\triangle \bar{\rho}^{(\alpha)}_{\bf p}$, which derive from the
field-dependent projection employed. (See the Appendix.)  
As a result, the projected Coulomb interaction
\begin{eqnarray}
\bar{H}^{\rm C}
={1\over{2}} \sum_{\bf p} \Big( V^{+}_{\bf p}
\bar{\rho}_{\bf- p}\,\bar{\rho}_{\bf p} +
4V^{-}_{\bf p} \bar{S}^{3}_{\bf- p}\, \bar{S}^{3}_{\bf p} \Big)
+ \triangle \bar{H}^{\rm C}
\label{projCoul}
\end{eqnarray}
acquires a field-dependent piece $\triangle \bar{H}^{\rm C}$, which
plays a crucial role, as we shall see.

The dynamics within the lowest Landau level is now governed by the
Hamiltonian $\bar{H}= \bar{H}^{\rm C} +\bar{H}^{\rm cyc} 
+\bar{H}^{\rm em}  + \bar{H}^{\rm tun}$.
Suppose now that an incompressible many-body state $|G\rangle$ of
uniform density $(\rho_{\rm av}^{(1)},\rho_{\rm av}^{(2)})$ is formed.
Then, setting  
$\langle G| \bar{\rho}^{(\alpha)}_{\bf -p} |G \rangle 
=\rho^{(\alpha)}_{\rm av}\, (2\pi)^{2}\delta^{2} ({\bf p})$ 
in $\bar{H}^{\rm em}$ one obtains the effective action to $O(A^{2})$:
\begin{eqnarray}
S^{\rm cycl}= -\int dtd^{2}{\bf x} 
\sum_{\alpha}\rho_{\rm av}^{(\alpha)}\,  {\cal U}^{(\alpha)}(x),
\label{Sem}
\end{eqnarray}
which summarizes the response due to electromagnetic
inter-Landau-level mixing, i.e., due to the cyclotron modes (one for
each layer). 

The electromagnetic interaction in $\bar{H}$ also gives rise to
intra-Landau-level transitions.
For single-layer systems the intra-Landau-level excitations  are only
dipole-inactive~\cite{GMP}  (i.e., the response vanishes faster than
${\bf k}^{2}$ for ${\bf k} \rightarrow 0$) as a result of Kohn's
theorem~\cite{Kohn},  
and the incompressible quantum Hall states show universal $O({\bf k})$
and $O({\bf k}^{2})$ long-wavelength electromagnetic characteristics
determined by the cyclotron mode alone.~\cite{KSdl}

The situation changes drastically for bilayer systems, 
where both in-phase and out-of-phase collective excitations arise.
In-phase excitations generally remain dipole-inactive, as a consequence
of invariance under translations of both layers.  Out-of-phase collective
excitations, in contrast, become dipole active~\cite{MZ,RR} (in the
absence of interlayer coherence) and modify the electromagnetic
characteristics of the bilayer systems substantially.~\cite{KSdl} 
Incompressible quantum Hall states well described by the Halperin
$(m,m,n)$ wave functions,~\cite{BIH} in particular, belong to this class
of states. The presence of interlayer coherence is expected to cause
further substantial changes in the systems, which we study in the
next section.

\section{Interlayer coherence and electromagnetic response}

In this section we study how the presence of interlayer coherence
affects the electromagnetic properties of bilayer systems.  
The particular set of states of our concern are the ground states at
filling $\nu =1/m$ for odd integers $m$, believed to have total
pseudospin $S=N_{e}/2$, with their orbital wave functions well
approximated by the Laughlin wave functions~\cite{L} or Halperin 
$(m,m,m)$ wave functions.~\cite{BIH} 
For definiteness we shall concentrate on the $\nu =1$ ground  state, 
but our analysis will apply to other cases as well.

Suppose first that the SU(2) breaking Coulomb interaction 
$V^{-}_{\bf p}= (e^{2}/4\epsilon)\, d + O(d^{2})$ is
negligibly weak (i.e., $d \rightarrow 0$ and 
$\triangle_{\rm SAS}\not=0$).
Then the $\nu =1$ ground state is given by the total-pseudospin
$S=N_{e}/2$ eigenstate $|G_{0}\rangle$,  fully polarized in the
$S^{1}$ direction via the tunneling interaction  so that 
$\la G_{0}|\bar{S}^{1}_{\bf p=0}|G_{0}\ra ={1\over{2}}\,N_{e}$, 
or
\begin{eqnarray}
&&\la G_{0}|\bar{S}^{a}_{\bf p}|G_{0}\ra 
=\delta^{a1}\,{1\over{2}}\,  \rho_{0}\, \delta_{\bf p,0},\ \ \ 
\la G_{0}|\bar{\rho}_{\bf p} |G_{0}\ra = \rho_{0}\, \delta_{\bf p,0},
\label{expvalue}
\end{eqnarray}
where $\rho_{0}= \rho^{(1)}_{\rm av} + \rho^{(2)}_{\rm av}$ and 
$\delta_{\bf p,0}\equiv (2\pi)^{2}\delta^{2}({\bf p})$.  
We suppose that this $S^{1}_{\bf p=0}=N_{e}/2$ eigenstate $|G_{0}\ra$
continues to be a good approximation to the $\nu =1$ ground state as
$V^{-}_{\bf p}$, kept weak, is turned on.  
It has been argued~\cite{Moon} and supported
experimentally~\cite{Murphy} that such $|G_{0}\ra$ well approximates
the ground state for $\triangle_{\rm SAS}\rightarrow 0$ with 
$V^{-}_{\bf p}\not=0$, where interlayer coherence 
$\la S^{1}\ra \not=0$ is realized spontaneously with $\la S^{3}\ra =0$
maintained so as to reduce the interlayer charging energy.

Further characterization of this $S^{1}_{{\bf p}=0}= N_{e}/2$
state is given by the static structure factors
\begin{eqnarray}
&&\la G_{0}|\bar{S}^{a'}_{\bf p} \bar{S}^{b'}_{\bf k}|G_{0}\ra 
= \delta_{\bf p+k,0}\,(\delta^{a'b'}\! + i\epsilon^{a'b'1})\,
{1\over{4}}\,\rho_{0}\, e^{-{1\over{2}}{\bf p}^{2}},\nonumber\\ 
&&\la G_{0}|\bar{\rho}_{\bf p} \bar{\rho}_{\bf k}|G_{0}\ra 
= \rho_{0}^{2}\, \delta_{\bf p,0}\, 
\delta_{\bf k,0} \equiv R_{\bf p,k}, \nonumber\\ 
&&\la G_{0}|\bar{\rho}_{\bf p} \bar{S}^{a}_{\bf k}|G_{0}\ra 
= 2\la G_{0}|\bar{S}^{1}_{\bf p} \bar{S}^{a}_{\bf k}|G_{0}\ra 
=\delta^{a1}\,{1\over{2}}\,R_{\bf p,k},
\label{projstr}
\end{eqnarray}
where $b'$ runs over $(2,3)$.
These relations are readily derived by rewriting $\bar{\rho}_{\bf p}$ and
$\bar{S}^{a}_{\bf p}$ in terms of the eigenspinors
$(\psi_{S},\psi_{A})$ of $S^{1}$,
and by noting that $|G_{0}\rangle$ involves no $\psi_{A}$
component (of $S^{1}=-{1\over{2}}$).   
For the partially-filled $\psi_{S}$ Landau level of $\nu = 1/3, 1/5, \cdots$ 
one has to retain in $R_{\bf p,k}$ a term~\cite{GMP} 
$\rho_{0}\,\delta_{\bf p+k,0}\, \bar{s}^{+}({\bf p})$ with
$\bar{s}^{+}({\bf p}) \sim O({\bf p}^{4})$, 
which vanishes in the present $\nu =1$ case.

The correlations characteristic of interlayer order 
$\bar{S}^{1}_{{\bf p}=0} = N_{e}/2$ are involved in
the structure factor
\begin{eqnarray} 
\bar{s}^{-}({\bf p})&=& {2\over{N_{e}}}\, \la G_{0}|\bar{S}^{3}_{\bf -p}
\bar{S}^{3}_{\bf p}|G_{0}\ra ={1\over{2}}\, e^{-{1\over{2}}{\bf p}^{2}},
\label{sminusk}
\end{eqnarray}
which is nonvanishing for ${\bf p} \rightarrow  0$, in contrast to
the case of the Halperin $(m,m,n)$ states 
where~\cite{RR,MZ} $\bar{s}^{-}({\bf p})\sim {\bf p}^{2}$.

Let us now study low-energy excitations over this
ground state.
With polarization $\la G_{0}|\bar{S}^{1}_{\bf p=0}|G_{0}\ra
= N_{e}/2$, the Coulomb interaction
$\bar{H}^{\rm Coul}$ has an (approximate) U(1) symmetry about the
$\bar{S}^{1}$ axis, yielding two Nambu-Goldstone (NG) modes 
$\{\Omega^{2}_{\bf p}(t),\Omega^{3}_{\bf p}(t) \}$. These NG modes
constitute the low-energy collective excitations in the system,
and one can  employ the technique~\cite{Moon} of nonlinear realizations
of the pseudospin symmetry for their description.   To this end let
$\Psi^{\rm cl}_{0}(y_{0},t)$ denote a classical configuration or the
ground-state configuration, characterized by the expectation values in
Eqs.~(\ref{expvalue}) and (\ref{projstr}).  Let us set 
$\Omega[{\bf r},t] \equiv \sum_{a}(\sigma^{a}/2)\,
\sum_{\bf p} \Omega^{a}_{\bf p}(t)\, e^{i{\bf p\cdot r}}$ (with
$a=2,3$) and write the electron field $\Psi_{0}$ in the form
of a small rotation in pseudospin from $\Psi^{\rm cl}_{0}$,
\begin{eqnarray}
\Psi_{0}(y_{0},t)=  e^{-i\Omega[{\bf r},t]}\,\Psi^{\rm cl}_{0}(y_{0},t).
\label{transformpsi}
\end{eqnarray} 
Here the NG modes serve as pseudospin textures in which the local
pseudospin alignment varies slowly with position. 
Rewriting the Lagrangian in favor of
$\Psi^{\rm cl}_{0}$ and $\Omega^{a}_{\bf p}$, and replacing the
products of $(\Psi^{\rm cl}_{0})^{\dag}$ and $\Psi^{\rm cl}_{0}$ by
the expectation values~(\ref{expvalue}) and (\ref{projstr})
then yields a low-energy effective Lagrangian for the NG modes
$(\Omega^{2},\Omega^{3})$.
  
To facilitate such transcription  it is convenient to
express~(\ref{transformpsi}) in operator form
\begin{eqnarray}
&&\Psi_{0}(y_{0},t) = {\cal P}\,\Psi^{\rm cl}_{0}(y_{0},t) {\cal P}^{-1},
\nonumber\\
&&{\cal P} = e^{i\Omega \cdot \bar{S}^{\rm cl}},\ \ \ 
\Omega \cdot \bar{S}^{\rm cl} \equiv \sum_{\bf p}
e^{{1\over{4}}{\bf p}^{2}}\Omega_{\bf p}^{a}  (\bar{S}^{\rm cl}_{\bf -p})^{a}.
\end{eqnarray} 
Here $(\bar{S}_{\bf p}^{\rm cl})^{a}$ stand for $\bar{S}_{\bf p}^{a}$ with
$\Psi_{0}$ replaced by $\Psi^{\rm cl}_{0}$, and obey the same
algebra~(\ref{chargealgebra}) as $\bar{S}_{\bf p}^{a}$. Repeated use of
the algebra then enables one to express  
$\bar{S}^{a}= {\cal P}\,(\bar{S}^{\rm cl})^{a}\, {\cal P}^{-1}$
and $\bar{\rho}^{a}= {\cal P}\,(\bar{\rho}^{\rm cl})\, {\cal P}^{-1}$
in powers of $\bar{\rho}^{\rm cl}$ and $(\bar{S}^{\rm cl})^{a}$.
[Remember that the characterization~(\ref{expvalue}) and
(\ref{projstr}) from now on applies to  
$(\bar{S}^{\rm cl}_{\bf p})^{a}$ and $\bar{\rho}^{\rm cl}_{\bf p}$.]
In particular, for their expectation values one obtains to
$O(\Omega^{2})$:
\begin{eqnarray}
\la \bar{\rho}_{\bf p}\ra 
&=& \rho_{0}\,\Big[\delta_{\bf p,0} +{1\over{2}}\, \gamma_{\bf p} 
\sum_{\bf k} \sin\Big({{\bf p}\!\times\! {\bf k}\over{2}} \Big)\,
\epsilon^{1a'b'}\Omega^{a'}_{\bf k}\,\Omega^{b'}_{\bf p-k}\Big],
\nonumber\\
\la \bar{S}^{1}_{\bf p}\ra
&=& {\rho_{0}\over{2}}\, \Big[ \delta_{\bf p,0}
-{1\over{2}}\, \gamma_{\bf p}\, \sum_{\bf k}\cos
\Big({{\bf p}\!\times\! {\bf k}\over{2}}\Big)
\Omega^{a'}_{\bf k}\, \Omega^{a'}_{\bf p-k}\Big] ,\nonumber\\
\la \bar{S}^{2}_{\bf p}\ra
&=& {\rho_{0}\over{2}}\, \gamma_{\bf p}\, \Omega^{3}_{\bf p} , 
\ \ \ 
\la \bar{S}^{3}_{\bf p}\ra
= -{\rho_{0}\over{2}}\, \gamma_{\bf p}\, \Omega^{2}_{\bf p}  ,
\label{inducedpspin}
\end{eqnarray}
where $\la \cdots\ra = \la G_{0}| \cdots|G_{0}\ra$ for short 
and $\gamma_{\bf p}=e^{-{1\over{4}}{\bf p}^{2}}$; $a'$ and $b'$ run 
over (2,3). These expressions suggest us to rename, following Moon et
al.~\cite{Moon},
$(m_{2})_{\bf p}= \Omega^{3}_{\bf p}$ and $(m_{3})_{\bf p}
= -\Omega^{2}_{\bf p}$ so that their $x$-space representatives
$m_{a}(x)=(m_{1}(x),m_{2}(x),m_{3}(x))$ stand for the pseudospin
density [with normalization $\sum_{a=1}^{3}(m_{a})^{2} \approx 1$ 
classically].  
Actually it is possible to generalize Eq.~(\ref{inducedpspin}) to all
powers of $m_{a}$, if one ignores their derivatives
$\partial_{k}m_{a}$:
\begin{eqnarray}
\la \bar{S}^{1}(x) \ra 
&\approx& {\rho_{0}\over{2}}\, \cos |{\bf m}|,\
\la \bar{S}^{2}(x) \ra
\approx {\rho_{0}\over{2}}\, {m_{2} \over{|{\bf m}|}}\,\sin |{\bf m}|,
\label{inducedpspinall}
\end{eqnarray}
where ${\bf m} = (m_{2}(x), m_{3}(x))$;
$\la \bar{\rho}(x)\ra \approx \rho_{0}$, etc.,

Moon {\it et al.}~\cite{Moon} earlier made such a pseudospin-texture
calculation and showed that the Coulomb interaction leads to the
following low-energy effective Hamiltonian to $O(\Omega^{2})$ and
$O({\bf p}^{2})$:
\begin{eqnarray}
 \la  \bar{H}^{\rm C} \ra
=\sum_{\bf p}\Big\{ &&\beta [{\bf p}]\,
|(m_{3})_{\bf p}|^{2} 
+{1\over{2}}\, \rho_{s}^{E}\,{\bf p}^{2}\,|(m_{2})_{\bf p}|^{2} 
\Big\},
\end{eqnarray}
with
\begin{eqnarray}
&&\rho_{s} = {1\over{8}}\, \rho_{0}\,
\sum_{\bf p} V^{11}_{\bf p}{\bf p}^{2} e^{-{1\over{2}}{\bf p}^{2}}
= {e^{2}\over{4\pi\epsilon \ell}}\,{\nu\over{16 \sqrt{2\pi}}}, 
\nonumber\\
&&\rho_{s}^{E}
= \rho_{s}\, \Big\{1 -\sqrt{8/\pi}\, \hat{d} + (3/2)\, \hat{d}^{2}
+ \cdots\Big\},\nonumber\\ 
&&\beta [{\bf p}] = \rho_{s}\,\Big\{(c_{0}/\ell^{2}) 
+ (c_{1}/\ell)|{\bf p}| +{1\over{2}}\, (1+ c_{2})\, {\bf p}^{2}\Big\},
\nonumber\\
&&c_{0}\approx \hat{d}^{2},\  c_{1} \approx -\sqrt{2/\pi}\,
\hat{d}^{2}, \   
c_{2} \approx  -\sqrt{8/\pi}\, \hat{d}\,(1 - \hat{d}^{2}/3) ,
\label{ICoul}
\end{eqnarray}
where we have used the pseudospin stiffness $\rho_{s}$ as a common
factor and recorded some corrections in powers of 
$\hat{d} \equiv d/\ell$; $\rho_{s}^{E}$ is given by the same 
expression as $\rho_{s}$ 
with $V^{11}_{\bf p}\rightarrow  V^{12}_{\bf p}$.

Substituting 
$\Psi_{0} =  e^{-i\Omega[{\bf r},t]}\,\Psi^{\rm cl}_{0}$
into the electronic kinetic term 
$\la \Psi^{\dag}_{0}\,i \partial_{t}\Psi_{0}\ra$ yields 
Berry's phase,~\cite{Berry} which turns into 
the kinetic term of the NG modes 
\begin{eqnarray}
 L^{\rm kin}
&=&  - {1\over{4}}\,\rho_{0}\,\sum_{\bf k}\epsilon^{ab1}
\Omega_{\bf -k}^{a} 
\partial_{t}\Omega_{\bf k}^{b} 
\end{eqnarray}
to $O(\Omega^{2})$. 
This shows that $\Omega^{2}=-m_{3}$ is canonically conjugate
to $\Omega^{3}= m_{2}$.

Substitution of Eqs.~(\ref{inducedpspin}) and~(\ref{inducedpspinall})
into $\bar{H}^{\rm em} + \bar{H}^{\rm tun}$ 
yields the coupling of the NG modes to external fields, 
$\la G_{0}|\bar{H}^{\rm em} + \bar{H}^{\rm tun}|G_{0}\ra 
= \int d^{2}{\bf x}\, {\cal H}_{A}$ with
\begin{eqnarray} 
{\cal H}_{A} =\rho_{0}\,\Big\{ && A^{+}_{0} 
 +{1\over{2}}\,\chi^{+}\,\epsilon^{ij}\,\partial_{i} m_{2}
\partial_{j}\,m_{3} \nonumber\\
&&+ \chi^{-}\,\gamma m_{3} 
-{1\over{2}}\,\triangle_{SAS}\,\cos |{\bf m}| \Big\} ,
\end{eqnarray}  
to $O(m^{2})$ and $O(\partial^{2)}$, 
where ${\bf m} = (m_{2}, m_{3})$, 
$\gamma = e^{{1\over{4}} \nabla^{2}}$ 
and $\chi^{\pm} =A_{0}^{\pm} + (1/2M)\, A_{12}^{\pm}$.

Similarly, the field-dependent Coulomb interaction 
$\triangle {\bar H}^{\rm C}$ leads to the effective interaction 
\begin{eqnarray} 
\la \triangle {\bar H}^{\rm C}\ra 
&=&  2\,\rho_{s}^{E}\int d^{2}{\bf x}\, \Big\{ 
m_{2}\,\partial_{j}A_{j}^{-}  + (A_{j}^{-})^{2} + \cdots \Big\};
\label{HdeltaH}
\end{eqnarray}  
see the Appendix for details.

Collecting terms so far obtained yields the effective action 
$S_{\rm eff}^{\rm coll}= \int dt d^{2}{\bf x}\, {\cal L}^{\rm coll}$ with
\begin{eqnarray}
{\cal L}^{\rm coll}
&=& {\rho_{0}\over{2}}\, m_{3}\, \Big(\dot{m}_{2} -2A_{0}^{-} \Big) 
- m_{3}\Big(\beta [{\bf p}] + {1\over{4}}\,\rho_{0}\triangle_{SAS}\Big)\,
m_{3}
 \nonumber\\ 
&&-{1\over{2}} \, \rho_{s}^{E} \,(\partial_{j}m_{2} -2 A_{j}^{-})^{2} 
+{1\over{2}}\,\rho_{0} \triangle_{SAS}\, \cos\, m_{2}
 \nonumber\\ 
&&-\rho_{0}\, A_{0}^{+}\Big( 1 +
{1\over{2}}\,\epsilon^{ij}\partial_{i}m_{2}\partial_{j}m_{3}\Big),
\label{Lcoll}
\end{eqnarray}
where ${\bf p} \rightarrow -i\nabla$ in $\beta [{\bf p}]$.
Here we have simplified the result slightly by retaining only terms
that contribute to the $O(\nabla^{2})$ electromagnetic response
eventually.  
The ${\cal L}^{\rm coll}$ is essentially the Lagrangian of a nonlinear
sigma model that supports classical topological 
excitations,~\cite{LK,SKKR} Skyrmions,  which constitute the low-lying
charged excitations of the system; see Eq.~(\ref{CPone}) in Sec.~V.  
Note that Eq.~(\ref{Lcoll}) correctly involves the topological charge
density~\cite{SKKR}
$(\rho_{0}/2)\,\epsilon^{ij}\partial_{i}m_{2}\partial_{j}m_{3}$,
which implies that the Skyrmions carry electric charge of a multiple 
of $\nu e$.

Let us here focus on the neutral collective excitations described 
by the field $m_{2}$ or $m_{3}$.
Eliminating $m_{3}$ from ${\cal L}^{\rm coll}$ yields the 
Lagrangian of the neutral field $m_{2}$ 
\begin{eqnarray}
{\cal L}^{\rm coll}_{m_{2}}
&=&  {1\over{2}}\, \rho_{s}^{E}\, \Big[
{1\over{v^{2}}}\,
(\partial_{t} m_{2} -2 {A'}^{-}_{\! 0})^{2} 
-(\partial_{j}m_{2} -2{A'}^{-}_{\!j})^{2} \Big] \nonumber\\
&&
+{1\over{2}}\,\rho_{0}\triangle_{SAS}\, \cos\, m_{2},
\label{Lcollmode}
\end{eqnarray} 
with 
\begin{eqnarray}
&&v^{2}= 2(\rho_{s}^{E}/\rho_{0}^{2})\, 
(4\beta [0] +\rho_{0}\triangle_{SAS}),\nonumber\\
&&2{A'}^{-}_{\! \mu} = 2 A^{-}_{\mu} -\partial_{\mu}\Gamma \approx d\,
(\partial_{z}A_{\mu} - \partial_{\mu}A_{z}).
\end{eqnarray}
Here we have indicated explicitly that $A_{\mu}^{-}$ so far used
actually stands for ${A'}^{-}_{\! \mu}$; we have also isolated the
$\rho _{0}A^{+}_{0}$ term which detects the charge of the ground state
$|G_{0}\ra$.

This collective mode $m_{2}$ gives rise to an electromagnetic response
of the form 
\begin{eqnarray}
{\cal L}^{\rm coll}_{\rm em}
&=&  2\rho_{s}^{E}\,
\Big(A_{j0}^{-}{\cal D} A_{j0}^{-}
- v^{2}A_{12}^{-}{\cal D}A_{12}^{-}\Big)  
\nonumber\\ &&
+\rho_{0}\,\triangle_{SAS}\,\Big( {A'}_{\!0}^{-}{\cal D} {A'}^{-}_{\!0} -
v^{2} {A'}^{-}_{\! j}{\cal D} {A'}^{-}_{\! j} \Big)
\label{emresonse}
\end{eqnarray} 
in compact notation, where ${\cal D}=1/\{\omega_{\bf p}^{2}
-(i\partial_{t})^{2}\}$ and 
\begin{eqnarray}
\omega_{\bf p}^{2}&=& \Big\{\triangle_{SAS} + {4\over{\rho_{0}}}\,
\beta [{\bf p}]\Big\} \Big\{\triangle_{SAS} +
{2\rho_{s}^{E}\over{\rho_{0}}}\, {\bf p}^{2}\Big\}
\label{omegap}
\end{eqnarray}
with ${\bf p} \rightarrow -i \nabla$. 
Here we have recovered $\beta [{\bf p}]$ to  obtain the dispersion
more accurately.
In terms of the field strengths in three space one can write 
${\cal L}^{\rm coll}_{\rm em}$ as
\begin{eqnarray}
{\cal L}^{\rm coll}_{\rm em}
&\approx&  {1\over{2}}\,\rho^{E}_{s} d^{2}\,
\Big(\partial_{z}{\bf E}_{\parallel}{\cal D}
\partial_{z}{\bf E}_{\parallel} -
v^{2}\,\partial_{z}B_{\perp}{\cal D}\partial_{z}B_{\perp}\Big)  
\nonumber\\ &&
+{\rho_{0}\over{4}}\,\triangle_{SAS}\,d^{2}\,\Big(E_{\perp}{\cal D} 
E_{\perp} 
- v^{2}\, {\bf B}_{\parallel} {\cal D}{\bf B}_{\parallel} 
\Big)
\end{eqnarray} 
in obvious notation.

The response due to the cyclotron modes in Eq.~(\ref{Sem}) is generally
suppressed by powers of $1/\omega_{c}$ compared with the collective-mode
contribution, except for the Hall-drift or Chern-Simons term 
\begin{eqnarray}
{\cal L}^{\rm cyc}_{A^{-}}
&=& 
- {\rho_{0}\over{2}} \, A^{-}_{\mu} \,
{\omega_{c}^{2}\over{ \omega _{c}^{2}} \!-\!
\omega^{2}}\,\epsilon^{\mu \nu\rho}
\partial_{\nu}A^{-}_{\rho} +\cdots,
\label{Halldrift}
\end{eqnarray}
which thus combines with  ${\cal L}^{\rm coll}_{\rm em}$ to form the
principal out-of-phase response of the system at long wavelengths. 
Note here that the collective mode gives rise to no such Hall-drift
term, unlike for the $(m,m,n)$ states.~\cite{KSdl}
This implies that no appreciable interlayer Hall drag is expected for the
present $\nu =1$ state, in contrast to the case~\cite{Renn} of the
(gapful) $(m,m,n)$ states.

Some comments are in order here.
First, the effective Lagrangian~(\ref{Lcollmode}) essentially
agrees with the one derived earlier~\cite{Moon,Etwo} 
if one sets $m_{2}=\hat{m}_{2} - \Gamma_{z}$, where $\hat{m}_{2}$ is
taken to undergo the gauge transformation 
$\delta \hat{m}_{2} = 2\theta^{-}$.
The earlier derivations focused on the spectrum of the low-lying
mode and its coupling to weak external electromagnetism was only 
guessed on the ground of gauge invariance.
A direct derivation of such electromagnetic coupling, as shown in our
approach, is quite nontrivial since it requires proper account of
Landau-level projection, especially the field-dependent Coulomb
interaction.

Second, in our approach electromagnetic gauge invariance is kept exact
at each step of discussion by use of the gauge-covariant fields 
$\psi'^{(a)}(x)$ and $A'^{-}_{\mu}(x)$ in Eq.~(\ref{transformedfield}). 
Recall that the pseudospin densities $\bar{S'}^{a}$ are
gauge-invariant while $S^{1}$ and $S^{2}$, defined in terms of
$\psi^{(a)}$, are gauge-variant so that  
$\bar{S}^{1} +i\bar{S}^{2} = e^{i\Gamma_{z}}\, 
(\bar{S'}^{1} +i\bar{S'}^{2})$.
Our characterization of interlayer coherence   
$\la G_{0}|\bar{S'}^{1}_{{\bf p}=0}|G_{0}\ra 
={1\over{2}} N_{e}$ therefore is a sensible gauge-invariant statement 
and, as a result, the related order parameters 
\begin{eqnarray} 
\langle \bar{S}^{1}_{{\bf p} =0}\rangle = 
{1\over{2}}\, N_{e}\, \cos\Gamma_{z}, \ \ \ 
\langle \bar{S}^{2}_{{\bf p} =0}\rangle =  
{1\over{2}}\, N_{e}\,  \sin\Gamma_{z},
\end{eqnarray}  
rotate in the pseudospin 1-2 plane under electromagnetic gauge
transformations $\Gamma_{z} \rightarrow \Gamma_{z} +2\theta^{-}$, 
or under the action of in-plane magnetic fields $\partial_{j}\Gamma_{z}$.
In other words, a naive choice  
$\langle \bar{S}^{a}_{{\bf p} =0}\rangle \propto \delta^{a1}$
is not physically acceptable unless layer spacing $d\rightarrow 0$.
This is the real reason why we have restarted with
$\psi'^{(a)}(x)$ and $A'^{-}_{\mu}(x)$ after Eq.~(\ref{transformedfield}).

We have handled two NG modes $(m_{2},m_{3})$ associated 
with SU(2) $\rightarrow$ U(1) breaking. 
They, being gauge-invariant, are neutral physical fields. 
They, however, happen to form a pair of canonical conjugates and thus
actually describe only one physical mode $m_{2}$. Note here
that, since $m_{2}\sim
\Omega^{3}$ , a shift $m_{2}
\rightarrow m_{2}+ {\rm const.}$ induces a rotation about the
$S^{3}$ axis so that
\begin{equation}
i[\bar{S}^{3}_{{\bf p}=0}, m_{2}] = 1 \not=0.
\end{equation}
This shows that $m_{2}$ can also be interpreted  as a NG mode
associated with the spontaneous breaking of the global U(1) symmetry
about the $S^{3}$ axis.~\cite{WZdlayer,EI} 
Because this global U(1)$^{-}$ is only approximate, $m_{2}$ is 
a pseudo NG mode and acquires a finite 
energy gap $\propto \triangle_{SAS}$. 
In the absence of tunneling 
($\triangle_{SAS}=0$ but $V^{-}_{\bf p}\not=0$), 
the U(1)$^{-}$ becomes exact but spontaneously broken; 
the energy gap closes and $m_{2}$ disperses linearly.

Unlike the global U(1)$^{-}$, the gauged U(1)$^{-}$ or U(1)$^{\rm em}$
is kept exact, as seen clearly from the gauge-invariant
response~(\ref{emresonse}).   This implies, in particular, that there
is no Anderson-Higgs mechanism or no Meissner effect working in the
present bilayer system.
Here we see a peculiar instance of spontaneous breaking of a global
symmetry with the related gauge symmetry kept exact; this
derives from the special character of the \lq\lq interlayer" gauge
invariance remarked in Sec.~II. 

Finally, one can use the effective theory to discuss the tunneling
phenomena.
The equation of motion of $m_{2}$ implies the conservation
law for the three-current $-\partial {\cal L}^{\rm coll}_{m_{2}}/\partial
A_{\mu}^{-} = j_{\mu}^{(1)}- j_{\mu}^{(2)}$, from which one can read off
the tunneling current 
$j^{\rm tun}_{z} \sim - \partial_{t}\rho^{(1)}$ as
\begin{eqnarray}
j^{\rm tun}_{z} = {1\over{2}}\,e\rho_{0}
\triangle_{SAS}\, \sin\, m_{2}.
\label{jtun}
\end{eqnarray}
Adding a source term $a_{z}\, j^{\rm tun}_{z}$ to
${\cal L}^{\rm coll}_{m_{2}}$ and calculating the response yields the
tunneling current 
\begin{eqnarray}
j^{\rm tun}_{z} =  {1\over{2}}\,e\rho_{0}
\triangle_{SAS}\, {1\over{ \omega^{2}} - \omega_{\bf p}^{2}}\,
\partial_{t}V_{z},
\end{eqnarray} 
in response to an alternating interlayer voltage 
$V_{z}=-2A'_{0}\approx -d\, E_{\perp}$.

\section{Relation to the single-mode approximation}

In this section we present a derivation of the electromagnetic
response~(\ref{emresonse}) by an alternative means, the single-mode
approximation (SMA).
Let us first suppose that $V^{-}_{\bf p}=0$, in which case 
the ground state is exactly given by 
the $\bar{S'}^{1}_{{\bf p}=0}= N_{e}/2$ eigenstate $|G_{0}\rangle$ in
Eq.~(\ref{expvalue}).

We consider the phonon-roton mode coupled to $A_{0}^{-}$ and represent
it as $|\phi^{-}_{\bf k}\rangle \sim \bar{S}^{3}_{\bf k}|G_{0}\rangle$.
The basic quantity in the SMA is the static structure factor
$\bar{s}^{-}({\bf k}) \sim 
\langle \phi^{-}_{\bf k}| \phi^{-}_{\bf k}\rangle$, 
which, in view of Eq. (\ref{sminusk}), is given by 
\begin{equation}
\bar{s}^{-}({\bf k})=(1/2)\, e^{-{1\over{2}}{\bf k}^{2}} .
\end{equation}

To determine the collective-excitation spectrum  
in the SMA one considers the(projected) oscillator strength
\begin{equation}
\bar{f}^{-}({\bf k}) = (2/N_{e})\, \langle G_{0}|\,
\bar{S}^{3}_{\bf -k}\, [\bar{H} , \bar{S}^{3}_{\bf k}]\,
|G_{0}\rangle,
\end{equation}
which is calculable~\cite{MZ,RR} by use of the
algebra~(\ref{chargealgebra}).
With $\bar{H}^{\rm tun} = - \triangle_{SAS}\,
\bar{S'}^{1}_{{\bf p}=0}$ included,
it is given to $O({\bf k}^{2})$ by 
\begin{equation}
\bar{f}^{-}({\bf k})
= {1\over{2}}\,e^{-{1\over{2}}{\bf k}^{2}}\, \Big[ \triangle_{SAS} +
2(\rho_{s}^{E}/\rho_{0})\, {\bf k}^{2} + \cdots\Big]. 
\label{fminus}
\end{equation}
Here the coefficient of the ${\bf k}^{2}$ term derives from the
general expression 
\begin{equation}
{1\over{2}}\,\sum_{\bf p} {\bf p}^{2}\ V^{12}_{\bf p}\,  
\{ \bar{s}^{-} ({\bf p}) -\bar{s}^{+} ({\bf p}) \}
\label{smimusplus}
\end{equation}
upon substitution of $\bar{s}^{-} ({\bf p})$ above; 
$\bar{s}^{+}({\bf p}) =0$ for $\nu=1$. 
Saturating $\bar{f}^{-}({\bf k})$  with the single mode 
$|\phi^{-}_{\bf k}\rangle$ then yields the  SMA excitation spectrum
$\epsilon^{-}_{\bf k}=\bar{f}^{-}({\bf k})/\bar{s}^{-}({\bf k})$ or
\begin{equation}
\epsilon^{-}_{\bf k}=\triangle_{SAS} 
+ 2(\rho_{s}^{E}/\rho_{0})\, {\bf k}^{2} +\cdots .
\label{eminus}
\end{equation}
This agrees with the spectrum derived by the pseudospin-texture
calculation in Eq.~(\ref{omegap}) with $V^{-}_{\bf p} \rightarrow 0$.

To calculate the electromagnetic response one may resort to the
previous SMA analysis,~\cite{KSdl} which, though developed originally
for the case of a dipole-active response
$\bar{s}^{-}({\bf k})=(c^{-}/2)\, {\bf k}^{2} +\cdots$, is
adapted to the present case as well:
One may simply replace $2\bar{s}^{-}({\bf k})\, \epsilon^{-}_{\bf k}$ 
in Eq.~(3.20) of Ref.~\onlinecite{KSdl} by 
$2\bar{f}^{-}({\bf k}) =\triangle_{SAS}\, e^{-{1\over{2}}{\bf k}^{2}} +
2(\rho_{s}^{E}/\rho_{0})\, {\bf k}^{2}$
and $c^{-}\epsilon_{0}^{-}$ in Eq.~(3.28) of Ref.~\onlinecite{KSdl} 
by Eq.~(\ref{smimusplus}) or $2 \rho_{s}^{E}/\rho_{0}$.
Then our result~(\ref{emresonse}) is correctly reproduced, apart from 
the $v^{2} A_{12}^{-} \cdots A_{12}^{-}$ and
$v^{2} {A'}_{j}^{-} \cdots {A'}_{j}^{-}$ terms, 
which, being $O(v^{2}) \sim O([\bar{H}^{\rm C}]^{2})$ higher in the
Coulomb interaction, were not covered in the previous SMA treatment.


It is possible to include the effect of $V^{-}_{\bf p}$ and make the
agreement complete if one appeals to the low-energy effective theory
in Eq.~(\ref{Lcoll}). With the identification 
$\bar{S}^{3}_{\bf p}= (\rho_{0}/2)\, \gamma_{\bf p}\, (m_{3})_{\bf p}$, 
as implied by Eq.~(\ref{inducedpspin}),
one can calculate $\bar{s}^{-}({\bf k})$ from the vacuum expectation
value $\propto (\gamma_{\bf k})^{2} \langle 0|(m_{3})_{\bf  -k}\, 
(m_{3})_{\bf k}|0\rangle$.
The task is thus reduced to determining the uncertainty 
$\langle 0|(m_{3})^{2} |0\rangle$ for a collection of harmonic
oscillators described by the Hamiltonian
\begin{eqnarray}
H^{\rm coll}
&\approx& \sum_{\bf k}\,
{1\over{4}}\,\left[g^{(12)}\,  |(m_{2})_{\bf k}|^{2}  
+ g^{(11)}\, |(m_{3})_{\bf k}|^{2} \right],
\label{Hsqueeze}
\end{eqnarray}
where $g^{(12)}\equiv \rho_{0}\triangle_{SAS} +
2\rho_{s}^{E} \, {\bf k}^{2}$ and  $g^{(11)}\equiv 
\rho_{0}\triangle_{SAS} + 4\, \beta [{\bf k}]$.  
Via rescaling $(m_{3})^{2} (\rho_{0}/2)\sqrt{g^{(11)}/g^{(12)}}$
is seen to attain the minimum uncertainty 
$(\hbar/2)\, \int d^{2}{\bf x}$, yielding
\begin{eqnarray}
\bar{s}^{-}({\bf k}) 
&=& {1\over{2}}\,  e^{-{1\over{2}}{\bf k}^{2}} 
\sqrt{g^{(12)}/g^{(11)}}.
\end{eqnarray}
The $\bar{f}^{-}({\bf k})$ in Eq.~(\ref{fminus}),
being already exact to $O(V^{-}_{\bf p})$, remains unmodified. 
The excitation spectrum and the response thereby agree with
those in Eq.~(\ref{emresonse}). 
The $\bar{s}^{-}({\bf k})$ above neatly summarizes the effect of
squeezing~\cite{NA} in pseudospin of the ground state due to two
competing sources of SU(2) breaking, $V^{-}_{\bf p}$ and
$\triangle_{SAS}$. 
It is seen from 
$\langle (m_{3})^{2}\rangle / \langle (m_{2})^{2}\rangle  =
g^{(12)}/g^{(11)} \propto (\bar{s}^{-}({\bf k}))^{2} $ 
that $\langle (m_{3})^{2}\rangle$ gets rapidly squeezed with 
decreasing $\triangle_{SAS}$, i.e., in passing from 
the tunneling regime to the correlation regime (where
$\bar{s}^{-}({\bf k}) \propto |{\bf k}|$ for $\triangle_{SAS}=0$).
It is an advantage of the pseudospin-texture theory that it
accommodates different types of correlations in a single framework.

\section{Comparison with the Chern-Simons approach}

In this section we examine the bilayer system within the Chern-Simons
theory. For the $\nu =1$ quantum Hall state, as naively described by
the $(1,1,1)$ state, one introduces a single CS field~\cite{EI} to
convert the electron fields $\psi'^{(\alpha)}$  [of
Eq.~(\ref{transformedfield})] into the composite-boson fields
$\psi^{(\alpha)}_{\rm cb}$.

Let us set $\psi^{(\alpha)}_{\rm cb}(x)=
\sqrt{\rho^{(\alpha)}(x)}\,e^{i\eta^{(\alpha)}(x)}$,
rewrite the Lagrangian in favor of 
$\rho^{\pm}= \rho^{(1)} \pm \rho^{(2)}$ and  
$\eta^{\pm} = \eta^{(1)} \pm \eta^{(2)}$, and expand it around the
mean-field $\rho^{+}(x) \sim \rho_{0}$. 
Then the $(\rho^{+}, \eta^{+})$ sector, coupled to $A_{\mu}^{+}$, is
seen to be  essentially the same as in the single-layer case. 
The $(\rho^{-},\eta^{-})$ sector, on the other hand, is sensitive to 
the SU(2) breaking interactions $\propto V^{-}_{\bf p}$ or 
$\triangle_{SAS}$.
Integration over $\rho^{-}$ leads to a low-energy Lagrangian,
that takes essentially the same form as  ${\cal L}^{\rm coll}_{m_{2}}$ 
in Eq.~(\ref{Lcollmode}) with $m_{2} \rightarrow \eta^{-}$, 
apart from some differences in scale.

The difference is subtle for the $(\partial_{0}m_{2} - 2A'_{0})^{2}$ term:
\begin{eqnarray}
&&v^{2}/\rho_{s}^{E}  \leftrightarrow  4\,V^{-}_{{\bf p}=0}
+ 2\triangle_{SAS}/\rho_{0} .
\label{coeffzero}
\end{eqnarray}
These coincide if $V^{-}_{{\bf p}=0}$ reads
$V^{-}_{{\bf p}=0} -(1/\rho_{0})\sum_{\bf p}V^{-}_{\bf p}e^{-{1\over{2}}\,
{\bf p}^{2}}$; this shows the importance of Landau-level projection, 
of which no explicit account is taken in the CS approach.
For the $(\partial_{j}m_{2} - 2A'_{j})^{2}$ term the discrepancy is 
\begin{eqnarray}
\rho_{s}^{E}  \leftrightarrow (\rho_{0}/4M)= \omega_{c}/(8\pi). 
\label{coeff}
\end{eqnarray}
Here we see that the CS approach attributes the pseudospin stiffness
improperly to inter-Landau-level processes.
Another difficulty is that an important Hall-drift
response~(\ref{Halldrift}) is missing from the CS theory.

All these subtleties derive from the fact that the CS approach,
because of the lack of Landau-level projection, fails to distinguish
between the cyclotron modes and the collective modes. 
The flux attachment in the CS approach properly introduces some crucial
correlations among electrons, but unfortunately not all of them.

Finally, it will be instructive to refer to the full effective
Lagrangian to make it clear what is missing.
Let us employ the decomposition~\cite{LK} 
$\psi^{(\alpha)}_{\rm cb}(x)=\sqrt{\rho(x)}\,Z^{\alpha}(x)$ in terms of 
a $CP^{1}$ field $Z=(Z^{1},Z^{2})^{\rm tr}$ with $Z^{\dag}Z=1$,
particularly suited for studying the dynamics of Skyrmions and vortices. 
Making use of the dual transformation of Lee and Zhang~\cite{LZ} then
enables one to rewrite the Lagrangian in terms of
$Z^{\alpha}$ and a vector field $b_{\mu}$ (representing the cyclotron
mode coupled to $A_{\mu}^{+}$):
\begin{eqnarray}
{\cal L}^{\rm CS}
&=& -( A^{B}_{\mu}\!+\!A_{\mu}^{+} -iZ^{\dag}D_{\mu}Z )
(\rho _{0}\delta_{0\mu}+\epsilon^{\mu \nu\rho}
\partial_{\nu}b_{\rho} )\nonumber\\
&&+ {\pi\over{\nu}}\, \left\{
b_{\mu}\epsilon^{\mu \nu\rho}
\partial_{\nu}b_{\rho}  + {1\over{\omega_{c}}}(b_{k0})^{2}\right\}
\nonumber\\
&& - {1\over{2}}K
\Big\{ |D_{k}Z^{\alpha}|^{2}+ (Z^{\dag}D_{k}Z)^{2} \Big\}
\nonumber\\
&& +{1\over{2}}\rho_{0}\,\triangle_{SAS}\,Z^{\dag}\sigma_{1}Z  + \cdots,
\label{CPone}
\end{eqnarray}
where only the principal terms are shown; 
$D_{\mu}=\partial_{\mu} +iA^{-}_{\mu}\sigma_{3}$ and
$b_{k0}=\partial_{k}b_{0} - \partial_{0}b_{k}$.
The last two terms, constituting a $CP^{1}$ nonlinear sigma model
with a breaking interaction, essentially coincide with 
our ${\cal L}^{\rm coll}$ in Eq.~(\ref{Lcoll})
if one replaces the stiffness  $K=\rho_{0}/M$ in this CS theory by 
$K=4 \rho_{s}$ [in accordance with Eq.~(\ref{coeff}) ] and 
includes some SU(2) breaking terms coming from $V^{-}_{\bf p}$.
The full effective Lagrangian is obtained by supplying to this
modified ${\cal L}^{\rm CS}$ the missing cyclotron-mode contribution
with another vector field $b_{\mu}^{-}$:
\begin{eqnarray}
{\cal L}^{-} &=& - A_{\mu}^{-} 
\epsilon^{\mu \nu\rho} \partial_{\nu}b^{-}_{\rho} \nonumber\\ 
&&
+ {\pi\over{\nu}} \left[ b^{-}_{\mu}\epsilon^{\mu \nu\rho}
\partial_{\nu}b^{-}_{\rho} 
+{1\over{\omega_{c}}}(b^{-}_{k0})^{2} \right] +\cdots.
\end{eqnarray}

\section{Summary and discussion}

In this paper we have studied the electromagnetic characteristics of
bilayer quantum Hall systems in the presence of interlayer coherence
and tunneling by means of a pseudospin-texture effective theory and
the single-mode approximation (SMA). It will be clear from the
analysis that a proper choice of the fields to start with, as well as
proper account of Landau-level mixing, is crucial for deriving a
long-wavelength effective theory in gauge-invariant form.  
We have seen from the response that electromagnetic gauge invariance
is kept exact, this, in particular, implying the absence of the
Anderson-Higgs mechanism or the Meissner effect in bilayer systems. 
The response also shows that no appreciable Hall drag is expected 
for the $\nu =1$ state, in contrast to the case of 
the gapful $(m,m,n)$ states. 
We have further seen that the identification of the low-lying neutral 
collective mode with a (pseudo) Nambu-Goldstone mode offers 
a peculiar instance of a spontaneously-broken (approximate) global 
symmetry with the related gauge symmetry kept intact. 
Our approach offers a critical look into the Chern-Simons theories, 
and we have observed that the lack of Landau-level projection is 
the principal source of subtleties inherent to them.

The idea underlying our approach is to explore the quantum Hall
systems via their electromagnetic response, which in some cases
is calculable without the details of the microscopic dynamics. 
An immediate example is the case of single-layer systems where it is
generally known that intra-Landau-level collective excitations are
dipole-inactive; the leading long-wavelength response of the
single-layer systems to $O({\bf k}^{2})$ therefore is governed by the
cyclotron mode alone. The second example is offered by bilayer systems
(without interlayer coherence), for which one can construct from the
response an effective gauge theory properly realizing the SMA spectrum
of collective excitations.  
The third example is the analysis of the effects of interlayer
coherence and tunneling given in the present paper.
These would combine to enforce again the fact that incompressibility is
the key character of the quantum Hall states and prove that studying the
response offers not only a fresh look at the quantum Hall systems
but also a practical means for constructing effective theories without
referring to composite bosons and fermions.

\acknowledgments

The author wishes to thank Z.~F.~Ezawa for useful discussions.
This work is supported in part by a Grant-in-Aid for Scientific
Research from the Ministry of Education of Japan, Science and Culture
(No. 14540261).

\appendix

\section{Field-dependent Coulomb interaction}
In this appendix we display some expressions related to 
the field-dependent Coulomb interaction $\triangle \bar{H}^{\rm C}$.    
The charge densities $\rho^{(\alpha)}_{\bf p}$ projected to the lowest
Landau level read 
$\bar{\rho}^{(\alpha)}_{\bf p} +
\triangle \bar{\rho}^{(\alpha)}_{\bf p}$ with 
\begin{eqnarray}
\triangle \bar{\rho}_{\bf p}^{(\alpha)}
&=& \sum_{\bf k} u^{(\alpha)}_{\bf p,k}\, \bar{\rho}_{\bf p-k}^{(\alpha)}
+\sum_{\bf q,k}\, w^{(\alpha)}_{\bf p,q,k}\, \bar{\rho}_{\bf p-q-k}^{(\alpha)}
+\cdots, 
\nonumber \\ 
u^{(\alpha)}_{\bf p,k}
&=& i\epsilon^{0jk}  p_{j}\,(A^{(\alpha)}_{k})_{\bf k}  + \cdots, 
\nonumber \\
w^{(\alpha)}_{\bf p,q,k}&=& 
-{1\over{4}}\,  {\bf p}^{2}\,\sum_{\bf q,k}\,
(A^{(\alpha)}_{i})_{\bf q}\,(A^{(\alpha)}_{i})_{\bf k}+ \cdots,
\label{deltarho}
\end{eqnarray}
where we have retained only terms with no derivatives acting on
$A^{(\alpha)}_{\mu}$, the portion relevant to
our discussion. 
They give rise to the field-dependent piece 
$\triangle \bar{H}^{\rm C}$ in the Coulomb interaction. 
See Ref. \onlinecite{KSdl} for the explicit form of the $O(A)$
contribution, which involves operator products of the form 
\begin{eqnarray}
&&\bar{I}^{+}_{\bf p,k} = \{\bar{\rho}_{\bf -p},\bar{\rho }_{\bf p-k}\},\ 
\bar{I}^{-}_{\bf p,k} = 2\{\bar{\rho}_{\bf -p},\bar{S}^{3}_{\bf p-k}\},
\label{dplusH}
\end{eqnarray}
and those with $\bar{\rho} \leftrightarrow 2\bar{S}^{3}$ in the above.

In Sec.~III we evaluate the expectation value 
$\langle \triangle \bar{H}^{\rm C}\rangle = \langle G_{0}|\triangle
\bar{H}^{\rm C}|G_{0}\rangle$ to derive an effective electromagnetic
coupling following from $\triangle \bar{H}^{\rm C}$. 
A direct calculation to $O(\Omega)$ shows that 
$\la \bar{I}^{-}_{\bf p,k} \ra = -\la \bar{I}^{-}_{\bf k-p,k} \ra
\propto \Omega^{3}_{\bf -k}$ 
while $\la \bar{I}^{+}_{\bf p,k}\ra \propto \delta_{{\bf k,0}}$
and $\la  \bar{S}^{3}_{\bf -p} \bar{S}_{\bf p-k}^{3}\ra
\propto \delta_{{\bf k,0}}$ fail to contribute. 
As a result, the $O(A)$  coupling is written as
\begin{eqnarray} 
\rho_{0}\,\sum_{\bf p,k} u^{-}_{\bf p,k}\,  V^{12}_{\bf p}
\gamma_{{\bf p}}\,\gamma_{{\bf k-p}}\,\sin \Big({{\bf p}\!\times\!
{\bf k}\over{2}}
\Big)\,\Omega^{3}_{\bf -k}.
\end{eqnarray}  
The calculation of the $O(A^{2})$ term is somewhat tedious, 
though straightforward, eventually leading to Eq.~(\ref{HdeltaH}).


\end{document}